\documentclass[oneside,a4paper,reqno]{amsart}
\usepackage{amssymb}

\begin{document}
\begin{titlepage}
\begin{center}
\bfseries  	PROBABILITIES ARE SINGLE-CASE, OR NOTHING
\end{center}
\vspace{1 cm}
\begin{center} D M APPLEBY
\end{center}
\begin{center} Department of Physics, Queen Mary
University of London,  Mile End Rd, London E1 4NS,
UK
 \end{center}
\vspace{0.5 cm}
\begin{center}
  (E-mail:  D.M.Appleby@qmul.ac.uk)
\end{center}
\vspace{0.75 cm}
\vspace{1.25 cm}
\begin{center}
\vspace{0.35 cm}
\parbox{12 cm }{ 
Physicists have, hitherto, mostly adopted a frequentist
conception of probability, according to which probability
statements apply only to ensembles.    It is argued that we
should, instead, adopt an epistemic, or Bayesian  conception,
in which probabilities are conceived as logical constructs
rather than physical realities, and in which probability
statements do apply directly to individual events.  The
question is closely related to the disagreement between the
orthodox school of statistical thought  and the Bayesian
school.  It has important technical implications (it makes a
difference, what statistical methodology one adopts).  It
may also have important implications for the interpretation
of the quantum state.
   }
\end{center}
\end{titlepage}
\section{Introduction}
\label{sec:Introduction}
The thoughts which follow were originally stimulated by some
conversations with Chris
Fuchs~\cite{FuchsSam,FuchsEtAl,FuchsSchack} concerning
probability, and the foundations of quantum mechanics.   These
conversations had a major impact on my thinking:  for they
caused me to see that the frequentist conception of
probability, which I had hitherto accepted, is deeply
confused.  This paper has grown out of my attempts to arrive at
a more satisfactory conception (also see
Appleby~\cite{Appleby}).

The first major applications of probability   to 
problems of theoretical physics were made by Laplace,
starting from the earlier work of Bayes.  Laplace took what
I will call an epistemic, or normative view of probability
(for an historical and conceptual overview of probability theory
see Gillies~\cite{GilliesA}, Hald~\cite{Hald},
Sklar~\cite{SklarA} and von Plato~\cite{vonPlato}). 

 Laplace was an uncompromising determinist.  He
considered that for ``an intelligence
sufficiently vast \dots nothing would be uncertain and the
future, as the past, would be present to its eyes''
(Laplace~\cite{Laplace}, p.4).  So for Laplace there could
be no question of probabilities existing objectively, out
there in the world, independently of ourselves.  Instead, he
regarded the theory of probability as what
Jaynes~\cite{JaynesA} calls ``extended logic'':  a process of
reasoning by which one extracts uncertain conclusions from
limited information.

The Laplacian view, of probability as logic, is well
described by Maxwell:
\begin{quote}
``The actual science of logic is conversant at present only
with things either certain, impossible, or entirely
doubtful, none of which (fortunately) we have to reason on. 
Therefore the true logic for this world is the calculus of
Probabilities, which takes account of the magnitude of the
probability which is, or ought to be, in a reasonable man's
mind'' (James Clerk Maxwell, quoted by Jaynes~\cite{JaynesA})
\end{quote}
On this view a probability statement is, not a
statement about what is \emph{in fact} the case, but a 
statement
about what one can \emph{reasonably expect} to be the case. 
Suppose that Alice buys one ticket in a lottery having $10^6$
tickets, and suppose that the ticket wins.  Then we can say:
\begin{itemize}
\item[(a)]
 Alice did in fact win the lottery.
\item[(b)]
Alice could not reasonably have expected to win the
lottery.
\end{itemize}
The second statement has a completely different logical
character from the first.  The first statement is a purely
factual statement concerning the lottery outcome.  The
second statement, by contrast, is a normative statement,
concerning the reasonableness of Alice's thoughts.

Laplace's epistemic interpretation of probability statements 
was, for a time, widely accepted.  However, in the latter
part of the
$19^{\mathrm{th}}$ century it began to go out of fashion,
and it has remained out of fashion ever since (although it
has continued to excite the interest of a small, though 
important
minority~\cite{JaynesA,JaynesB,JeffreysB,JeffreysC,Ramsey,Fin2,Fin1,Savage,Bernardo,Howson,Earman}). 
Instead the vast majority of
$20^{\mathrm{th}}$ century scientists, mathematicians and
philosophers have favoured what, for want of a better term,
I am going to call an objectivist interpretation: either a
frequency
interpretation~\cite{Mises,MisesB,Reich,PopperB,Fraassen}, or
(in more recent years) a propensity 
interpretation~\cite{GilliesA,GilliesB,PopperC,PopperD}.  

This change in conceptual standpoint led to major changes in
statistical methodology.  Under the influence of objectivist
ideas Fisher and others~\cite{Hald,Fisher} rejected the Bayesian
methodology favoured by Laplace, and developed in its place
what is now the orthodox methodology, described in every
textbook.  
  
Physicists are apt to ignore philosophical disputes on the
grounds that they are scientifically inconsequential.  This,
however, is a case where a dispute regarding the conceptual
fundaments has important practical implications, for just
about every area of scientific activity.  In particular, it
has important implications for the problem of quantum state
determination.  

It is easy to see what motivated the turn from epistemic to
objectivist.   Objectivists attempt to represent
probabilities as physically real quantities, similar to
quantities like mass or length, existing out there in the
world, wholly independent of us.    The attractions of this
programme to a physical scientist are obvious.  
Physical scientists are, by education, focussed outwards, on
the pursuit of mind-independent, empirical truth.   They
naturally rely on the internally derived suggestions of
hunch and intuition.  However, these suggestions are then
subjected to rigorous empirical testing.  To anyone with
this mindset the epistemic point of view is likely to seem
deeply unattractive:  for it means that there is a component
to probability which is fundamentally non-empirical.  

Unfortunately, the objectivist programme,  however  laudable
its motives, fails in its purpose.  Objectivist
interpretations do not really relieve us of the need to make
normative judgments.  They only disguise the fact
that we are making normative judgments.  Objectivists
dislike the Laplacian approach because they think it
subjective.   But their own approach is no less
subjective.  It is just that
they have found a way of making the fact less obvious.

The question at issue is closely related to the so-called  
problem of induction. Hume~\cite{HumeA}, more than 250 years
ago, noted that our expectation that the sun will rise tomorrow
cannot be based \emph{solely} on empirical facts concerning the
past behaviour of the solar system. Not only do those facts,
taken by themselves, not make it certain that the Earth's
angular momentum will continue to be approximately
conserved. They do not  even make it likely. If,
nevertheless, we confidently expect the angular momentum to
be conserved it is because we are tacitly supplementing our
past observations with an additional normative principle,
which says (roughly speaking) that regularities observed on
the past light cone may, subject to certain restrictions,
justifiably be extrapolated to events at space-like and
future time-like separations.  Hume's point was that this
principle cannot \emph{itself} be 
inferred from the observations.  

NASA engineers are so confident of the Newtonian law of
gravity that they are willing to stake the lives of their
astronauts on its being correct (to a good approximation,
within its domain of application).   Yet the data set on
which their confidence is based is strictly finite.   Viewed
\emph{sub specie aeternitatis}
 it is completely negligible.  Which
raises the question:  exactly how large does the data set
have to be in order for that degree of confidence to be
justified?  This is not itself a question which can be
settled experimentally.

Newton himself was concerned by this question, as appears
from the following:
\begin{quote}
``although the arguing from Experiments and Observations by
Induction be no Demonstration of general Conclusions; yet it
is the best way of arguing which the Nature of Things admits
of'' (Newton~\cite{Newton}, p.404)
\end{quote}
In other words:  we do not reason inductively because we
like it, but because we do not have a choice. 

Since then the question has attracted the attention of
numerous philosophers.  Popper~\cite{PopperB}, in particular,
has argued that science does not, in fact, rely on inductive
reasoning (for the opposite view see, for example,
Jaynes~\cite{JaynesA}, Stove~\cite{Stove} and
Newton-Smith~\cite{NewSmith}).  However, nothing that Popper or
anyone else has written changes the fact that
\begin{itemize}
\item[(a)]
 We mostly do believe that NASA's reliance on the
Newtonian law of gravity is reasonable, given the data.
\end{itemize}
and
\begin{itemize}
\item[(b)]
This normative belief is not itself based on empirical
facts about the solar system. 
\end{itemize}
Furthermore, if we really did refuse to entertain such
normative beliefs---if we really did think that scientific
predictions are no more reasonable than astrological
ones---then science would lose its point, at least so far as
its practical applications are concerned.  Science is
intellectually demanding.  It is laborious and
time-consuming.  If they did not think that the predictions
which result are normatively preferable, practically-minded
people would resort to some less demanding procedure, such
as  taking a blind guess.

Now it seems to me that induction is really just a special case
of probabilistic reasoning, and that what goes for the one goes
for the other too.  There is a strong family resemblance
between (a) the prediction that the Earth's angular momentum
will  still be conserved tomorrow and (b) the prediction 
that a coin will probably still come up heads approximately
$50\%$ of the time tomorrow.  If (a) does not follow just from
the observations, without additional assumptions, then nor, one
might suppose, does (b). 

 However,  $20^\mathrm{th}$ century
thought on the subject (both scientific and philosophical)
has, on the whole, been strongly resistant to that
suggestion.   It is widely accepted that induction involves a
normative assumption (and is therefore suspect, in the view of 
many). 
 But there has been a marked reluctance to
accept that the same is true of probability statements. 
 It seems to me that there is an
inconsistency here.

\section{Frequentism (1):  Infinite Ensembles}

Frequentism is the position that a probability statement is
equivalent to a frequency statement about some suitably
chosen ensemble.  For instance, according to von
Mises~\cite{Mises,MisesB} the statement ``the probability of
this coin coming up heads is
$0.5$'' is equivalent to the statement ``in an infinite
sequence of tosses this coin will  come up heads with
limiting relative frequency $0.5$''.  Of course, infinite
sequences are unobservable.   Consequently Popper~\cite{PopperB}
proposes the weaker position, that we regard the statement as
``methodologically falsified'' if, in a finite sequence of
$N$ tosses, the relative frequency of heads differs from
$0.5$ by more than $\epsilon$ (where $N$ is suitably large,
and $\epsilon$ is suitably small).  Other variants of the
basic frequentist idea are possible.

At first sight this idea may seem very plausible:  for the
notions of probability and frequency are obviously very
closely related.  If the reader does find it plausible s/he
should reflect that the frequentist position is not simply
that the notions of probability and frequency are intimately
\emph{connected}, but that they are actually
\emph{identical}.

From a mathematical point of view the cleanest way of trying
to implement the frequentist idea is to identify
probabilities with relative frequencies in infinite
ensembles.  However,  there is an obvious problem with that: 
for it is not very plausible to suppose that a given coin ever
will be tossed an infinite number of times---and if the universe
has finite $4$-volume it is definitely impossible.  von Mises
attempts to circumvent this difficulty by defining the
probability in terms of the limiting relative frequency
which \emph{would} be obtained if the coin 
counter-factually \emph{were}
tossed an infinite number of times.  In other words, von
Mises defines the probability, not in terms of an actually
existing ensemble out there in the world, but in terms of a
completely fictitious entity which does not exist anywhere
(not even in our  imagination).   As Jeffrey~\cite{Jeffrey}
says, that is not consistent with the idea that a probability 
is an objectively real quantity, similar to a mass.

This is not the only problem.  Suppose, for the
sake of argument, that the universe contains an infinite
space-like slice $S$ on which there are infinitely many 
$\mathstrut^{226}\mathrm{Ra}$
nuclei.  Let $A$ be the spherical region centred on the Earth
with (say) radius $10^{100}$ light years, and let $B$ be the
part of $S$ outside $A$.  Suppose that half the nuclei in $A$ decay
within proper time $1,600$ years, 
whereas half the nuclei in $B$
decay within proper time $1$ second (the second statement is,
of course, relative to the way one takes the limit for the
infinite collection of nuclei in $B$). Then, on an infinite
frequentist defintion, we would have to say that the
\emph{true} half-life is $1$ second, as defined by the
infinite ensemble consisting of all the nuclei on $S$. 
However, the  physically relevant
half-life (relevant, that is, to a physicist on Earth) is
$1,600$ years, as defined by the finite ensemble consisting
of all the nuclei in $A$.  Probabilities in the sense of an
infinite ensemble definition may conceivably exist.  But 
there is no reason to assume any connection between them and
the empirically important probabilities that interest us as
Earth-bound, experimental scientists.

This example is not fanciful.  There is nothing 
implausible in the suggestion that the standard model
parameters, which determine the half-life, may vary
significantly over distances $\sim 10^{100}$ light years.

The objection is quite general.  Suppose, for the sake of
argument, that a coin actually could be tossed an infinite
number of times.  Suppose that, in the first
$10^{10^{100}}$  years of its existence, the coin comes up
heads with relative frequency $0.5$, but that in the rest of
its infinite history it comes up heads with limiting
relative frequency $0.25$.  Then the empirically important
probability of heads (the probability that matters to 
\emph{us}) is
$0.5$, not $0.25$ as on an infinite frequentist definition.

An infinite frequentist must admit that it is logically
possible that a coin could come up heads with relative
frequency
$0.5$ in the first $10^{10^{100}}$  years of its existence,
and then with limiting relative frequency 0.25 thereafter. 
But s/he probably has, at the back of his or her mind, some
notion that it is not very \emph{likely}.  
That, however,  involves
a tacit appeal to assumptions which, on frequentist
principles, are  inadmissible.

Suppose that, instead of considering a sequence of coin
tosses, we considered the sequence which consists of $N$
tosses of a coin, followed by infinitely many tosses of a
die, with the convention that $1$ on the die counts as
``heads'', while $2$ or greater counts as ``tails''.   I do
not think that anyone would be inclined to infer, from the
fact that ``heads'' occurs with relative frequency $0.5$ in
the first $N$ tosses, that it is therefore likely to occur
with  limiting relative frequency $0.5$ in the rest of the
infinite sequence. If we are inclined to make that inference
in the case of the sequence which consists of infinitely many
tosses of a single coin, it is because in this case, but not
in the other, it seems natural to assume that the
probability of heads remains constant.  However, a
frequentist cannot consistently make that assumption.

If one assumes that the probabilities remain constant, that
means one is tacitly relying on some kind of notion of a
single-case probability   (the probability
cannot be the \emph{same} on every toss if it is not
\emph{defined} on every toss).  But on a frequentist
interpretation it makes no sense to speak of a single-case
probability.  On a frequentist interpretation probabilities
are properties of the whole ensemble, not of the individual
events (the probability \emph{simply is} the limiting
relative frequency).

But even if we were to allow this tacit
appeal to single-case probabilities, it would not make the
infinite ensemble definition any more acceptable.  The
behaviour of the propensity over infinitely great expanses
of time and/or space has absolutely no bearing on the
probabilities of immediate experimental interest. The
propensity of a coin for coming up heads may remain constant
over times greater than $10^{10^{100}}$  years (supposing
that the coin, not to mention the universe, lasts that
long).  Or it may not (the coin may become bent).  Either
way, it does not matter.  So far as the empirical
applications are concerned, all that matters is that the
propensity should remain constant over the strictly finite
patch of $4$-space in which we happen to be empirically
interested.   It follows that, if the frequentist approach
is to work at all, the definition had better be in terms of
finite ensembles.

\section{Frequentism (2):  Finite Ensembles}
\label{sec:FiniteFrequentism}
 The shift to
finite ensembles necessitates a significant weakening of the
definition.   In a finite number of independent
tosses, every sequence of heads and tails has probability
$ >0$
(unless the probability of heads = $0$ or $1$).  It follows that
probability statements cannot be \emph{strictly} 
equivalent to
statements about frequencies in finite ensembles.

The usual response to this difficulty is to argue that
probabilities sufficiently close to $0$ count as effective
impossibilities, and probabilities sufficiently close to $1$
count as effective certainties.  Consequently, the
proposition ``the probability of heads is $p$'', though not
strictly equivalent, is 
equivalent\footnote{
  Popper~\cite{PopperB} and Gillies~\cite{GilliesA} are more
cautious:  they only speak of 
  probability statements being FAPP falsified
  (``methodologically falsified'' in their terminology),
  never of them being FAPP confirmed. 
}   
FAPP (``for all practical purposes'') to the
proposition ``the relative frequency of heads will be very
close to $p$ in a sufficiently long sequence of tosses''.

On first inspection this idea may seem very persuasive.   It
seems, on the face of it, to accurately describe the way we
use probabilities in physics (we are, for example,
accustomed to think of the second law of thermodynamics as
FAPP deterministic in its application to macroscopic
systems).  Orthodox statisticians encourage us to think that
the principle ``highly improbable = FAPP impossible'' also
underpins the theory of statistical inference
(Fisher~\cite{FisherStat} (pp.~40--9), for example,
suggests\footnote{
  Perhaps I should say he \emph{seems} to suggest.  
In his actual practice he departs from this
principle---see below.
}  
that statistical arguments are based on
the ``resistance felt by the normal mind to accepting a
story intrinsically too improbable'').

However, I think it becomes clear on closer examination that
things do not  work quite in that way.   Finite ensembles
are, of course, an essential part of the empirical
interface:  the procedures we use for testing probability
assignments.  So probabilities are intimately
\emph{connected}  to frequencies in
finite ensembles.  However, the connection is not a FAPP
equivalence.  It is more subtle than that.

If a coin is tossed $100$ times then, on the hypothesis that
the coin is fair and the tosses independent, every possible
sequence of heads and tails has probability $2^{-100}$. On
this hypothesis the outcome is \emph{guaranteed} to be highly
improbable.  So if we really were relying on the principle
``highly improbable = FAPP impossible'' we would not only
take the hypothesis to be FAPP falsified by the sequence
consisting of $100$ heads.  We would take it to be just as
strongly falsified by (for example) a sequence consisting of
$50$ heads and $50$ tails in some ostensibly random order.  
There would, in fact, be no need to toss the coin at all: 
we would know in advance that the hypothesis was going to be
FAPP falsified, whatever  the outcome.

Highly improbable events do \emph{not} necessarily count as FAPP
impossible~\cite{Howson}. The probability of the microstate of
the air in the room where I am now writing is no greater than
the probability of observing a macroscopic violation of the
second law of thermodynamics.  Yet the fact that this
microstate  occurred is not occasion for surprise.

Highly improbable events are happening all the time.  It is,
of course, true that the occurrence of \emph{some} events, which we
previously took to be highly improbable, forces a revision
of our starting assumptions.  But it is also true that the
occurrence of many other events, which we previously
regarded as no less improbable, leaves our starting
assumptions intact.  The question is:  what distinguishes
the small class of improbable events, which do force a
revision, from the much larger class, which do not?

Consider
\begin{quote}
\emph{Argument A:}  Alice spins a roulette wheel once, and obtains
the number $11$.  She concludes that the wheel is fair.
\end{quote}
This argument seems clearly invalid.  Merely from the fact
that $11$ occurred once, one cannot reasonably infer that the
other numbers are even possible, much less that they all
have probability $1/37$.  It also seems a little strange to
argue that, because $11$ \emph{did} occur, therefore $11$ is rather
unlikely to occur (an anti-inductive argument, as it might
be called).

Now compare

\begin{quote}
\emph{Argument B:}  Bob tosses a coin $100$ times, and obtains a
sequence consisting of $50$ heads and $50$ tails in some
ostensibly random order.  He concludes that the coin is fair
(more precisely:  he is $95\%$ confident that the probability
of heads is in the interval $(0.4, 0.6)$).
\end{quote}
It may appear that this argument is, by contrast, valid. 
Yet if Bob really is relying purely on the observed facts,
and nothing else whatever, his argument is no better than
Alice's.

A sequence of $100$ coin tosses is, from the point of view of
abstract probability theory, equivalent to $1$ spin of a big
roulette wheel, divided into $2^{100}$ sectors.  Let
$\mathbf{s}$ be the particular sequence which Bob obtains.
Then, on the basis of one spin of the equivalent roulette
wheel, Bob is arguing 
\begin{itemize}
\item[(a)]
 Because  $\mathbf{s}$ \emph{did} occur, therefore
each of the sequences which \emph{did not} occur has
probability $\sim 10^{-32}$.
\item[(b)]
 Because $\mathbf{s}$ \emph{did} occur, therefore
$\mathbf{s}$ is \emph{very unlikely} to occur.
\end{itemize}
As it stands Bob's argument has the same extraordinary
features as Alice's.  So it is no more valid than
hers.

Of course, if a coin did in fact come up heads on $50$ out of
$100$ successive tosses, we mostly would conclude that the
probability of heads is close to $0.5$.  I am not suggesting
we would be wrong to make that inference.  However, we would
be basing ourselves, not merely on the observed facts, but
also on certain prior probabilistic assumptions.

Actually, the conclusion to Alice's argument would become
valid if she was allowed to make some additional
assumptions.  Suppose, for example, Alice knows there are
two types of roulette wheel on the market:  a wheel made by
a reputable manufacturer, which can safely be assumed to be
fair, and a trick wheel which always stops at the number
$10$.  Then the fact that Alice's wheel stops at $11$ shows that
it is not a trick wheel.  So she can validly infer that the
wheel is fair, just on the basis of spinning it once.

In the original statement of the problem Alice wants to
select one distribution out of the set of \emph{all possible}
distributions on the set of integers $0$Ð-$36$.  When the
choice is as wide as that, a single trial says very
little.   But when the choice is narrowed down , so that she
only has to choose between the uniform distribution, and the
distribution concentrated on the number $10$,  a single
trial can settle the issue.  \emph{Given her assumptions} Alice
knows with certainty that the distribution is uniform,
merely from spinning the wheel once and getting
the number $11$.

Of course, it is not usual for a statistical inference to
decide the question with certainty.  The following example
is closer to the situations one commonly meets in practice. 
Suppose it is possible to buy a certain type of random
number generator.   The machine has a button on the front,
and a display.  Pressing the button causes a randomly
selected $20$-digit integer to appear in the display.  If the
machine is working properly each integer $n$ in the range 
$0\le n < 10^{20}$  
has probability $10^{-20}$.  However, it is known that the
manufacturer put out a faulty batch.  In the faulty machines
each $n$ in the range $0\le n < 10^{10}$ has probability
$10^{-10}$,
while every other value has probability $0$.   Suppose, now,
that  we buy one of these machines, press the button, and
the number $00000000005678435211$ appears in the display. 
Those ten leading zeros may cause  us to suspect
that the machine comes from the faulty batch.  

The point to notice here is that we would clearly not be
relying on the principle ``highly improbable = FAPP
impossible''. Let $H_1$ be the hypothesis ``machine is
functioning correctly'', let $H_2$ be the hypothesis
``machine comes from the faulty batch'' and let $E$ be the
event ``number observed is $5,678,435,211$''.   Then
\begin{align}
P\left( E \left| H_1\right.\right) & = 10^{-20} \\
P\left( E \left| H_2\right.\right) & = 10^{-10}
\end{align}
 $E$ is highly improbable on either hypothesis.  
 The inference
is not based on the principle, that one should reject any
hypothesis which make the observation highly improbable, but
rather on the principle, that one should prefer the
hypothesis which makes it least improbable (at least, that
is the principle on which Fisher's likelihood method
relies; Bayesian statisticians introduce an important
modification---see below).

The absolute values of the conditional probabilities are
completely irrelevant.  The inference would go through just
the same if the conditional probabilities were, instead,
$P\left( E\left| H_1 \right. \right)=10^{-10^{20}}$ and
$P\left( E\left| H_1 \right. \right)=10^{-10^{10}}$ .

If we really were motivated by Fisher's supposed ``resistance
felt by the normal mind to accepting a story intrinsically
too improbable'' then we would reject \emph{both} of the 
stated
options, and choose instead a third, more congenial
hypothesis.   There is, for instance, no empirical
consideration which excludes the hypothesis that the number
$5,678,435,211$ was  fully determined by the state of the
machine before the button was pressed, so that
$P\left(E\left|H\right.\right)=1$. Indeed, if the machine
is really a $pseudo$-random number (as is likely) that
would, in fact, be the case.

Now the above are both instances of an inference based on a
single trial.  It may be thought that the inference in
argument $B$, being based on many repeated trials, would have
a completely different logic.  But that is not so.  There is
no fundamental difference between  inferences based on
singular events, and inferences based on large ensembles
(although there are, of course, some important differences
in point of detail).

In the examples just discussed the inference involved making
a choice between two competing hypotheses.  The inference in
argument $B$ is based on the same principle, except that now
the choice is between the non-denumerable infinity of
hypotheses
\begin{quote}
$H_p$ = ``the tosses are independent and the probability of
heads is $p$ on every toss''
\end{quote}
where $p$ ranges over the closed interval $[0,1]$.  Let $E$
 be the
observed outcome ``$50$ heads and $50$ tails in some ostensibly
random order''.  Then $P\left(E\left| H_p\right.\right) =
p^{50} (1-p)^{50}$.   We prefer hypothesis $H_{0.5}$ to (say)
$H_{0.2}$ because, although $P\left(E\left|
H_{0.5}\right.\right)=8\times 10^{-31}$  is very small, 
$P\left(E\left|
H_{0.2}\right.\right)=2\times 10^{-40}$  is
$9$ orders of magnitude smaller.  The fact that $E$,
 in \emph{absolute}
terms, is highly improbable on either hypothesis has nothing
whatever to do with it.

This restriction of the set of admissible hypotheses to the
$1$-parameter family $H_p$ is essential.  Without that, or
some other such restriction, no useful inference is
possible, as we saw in our earlier discussion of argument $B$.

The standard way of relating a probability to the frequency
observed in a sequence of repeated trials is thus critically
dependent on the assumptions that (a) the trials are
independent and (b) the probability is constant. 
We are so accustomed to making these assumptions in
theoretical calculations that they may appear trivial.  But
if one looks at them from the point of view of a warking
statistician it will be seen that they are very far from
trivial. 

The probability of a coin coming up heads depends as much on
the tossing procedure as it does on properties of the coin.
Suppose that, in an experiment to determine the probability,
one used a number of visibly different tossing procedures,
without keeping any record of which procedure was employed
on which particular toss.  We would mostly consider the
results of this experiment to be meaningless,  on the
grounds that the probability of heads might be varying in an
uncontrolled manner.  It is clearly essential, in any
serious experiment, to standardize the tossing procedure in
such a way as to ensure that the probability of heads is
constant.  This raises the question:  how can we be sure
that we have standardized properly?  And, more
fundamentally:  what does it
\emph{mean} to say that the probability is constant?  Anyone
who thinks these questions are easily answered  should
read chapter 10 of Jaynes~\cite{JaynesA}  (also see
Appleby~\cite{Appleby}, where I approach the question from a
different angle).

This problem, and problems related to it, has to be faced in
just about every statistical application.   For instance, the
 concept of an opinion poll, viewed in abstract
mathematical terms, is very simple.  What makes opinion
polling difficult in practice is (among other things) the
fact that it is, in practice, very hard to select
the sample in such a way that the trials are independent,
and  the probability of each individual respondent being a
supporter of party $X$ is constant, equal to the proportion
of
$X$-supporters  in the population as a whole.

Frequentists are impressed by the
fact that  we infer probabilities from
frequencies observed in finite ensembles. 
  What they overlook is the fact that we do not infer
probabilities from just
\emph{any} ensemble, but only from certain very carefully
selected ensembles in which the probabilities are, we
suppose, constant (or, at any rate, varying in a specified
manner).  This means that statistical reasoning makes an
essential appeal to   the concept of a single-case
probability:  for you cannot say that the probability is the
\emph{same} on  every trial if you do not accept that the
probability is \emph{defined} on every trial.  
The only question is whether the single-case probabilities
are to be construed as objective realities
(``propensities''), or whether they should be construed in
an epistemic sense.

\section{Bayesian Analysis}

As is well known,  gamblers are prone to think that, if a
coin has come up heads on each of (say) the last $5$ tosses,
it is more likely than not to come up tails on the next
toss.  One of the first things students are taught is that
this is a fallacy.  If the tosses are truly independent, and
if the coin is truly fair, then the probability of heads on
the next toss is still $0.5$, even  if the coin has come up
heads on each of the last $10^3$ tosses.

Of course, if a coin did, in practice, come up heads $10^3$
times in succession hardly anyone would stick to the belief
that the probability of heads is $0.5$.  But it is important
to realize that there is nothing in the data itself which
forces us to that conclusion.  A sequence of $10^3$ heads is
\emph{in itself} no more inconsistent with the hypothesis that the
coin is fair than a sequence of $500$ heads and $500$ 
tails in
some ostensibly random order (if it were, it would mean that
generations of students have been wrongly taught).   Our
decision to embrace the alternative hypothesis, that the
coin is biased, is a function, not of the data alone, but of
the data in combination with our subjective
 \emph{willingness} to
embrace the alternative hypothesis.  And that is
something which can vary.  The person who is absolutely certain
that the coin is fair, and who holds to that belief even though
the coin has come up heads on each of the last $10^3$ tosses,
is not guilty of any inconsistency (not even an inconsistency
FAPP, as elementary textbooks correctly emphasize).

This raises the question:  just how willing should one be? 
Alice is very open-minded:  $10$ heads in succession is enough
to convince her that the coin is probably biased.  Bob is
harder to persuade:  it takes $50$ heads in succession to
convince him.  Which of them is right?

Let us go back to the random number generator we considered
in the last section.   I suggested that if one presses the
button once and the number $5,678,435,211$ appears in the
display, then there is reason to suspect that the machine is
faulty.    There is a problem with that, however.   Suppose,
for instance, that the manufacturer constructed $10^{20}$ of
these machines, and only one of them was faulty (this is
not, of course, a very plausible assumption, but let us make
it anyway).  In that case the improbability of getting the
number $5,678,435,211$ on the hypothesis that the machine is
working properly is outweighed
by the much greater improbability, that we should have
chanced to buy that single faulty one.  So the inference
would not be justified.  If, on the other hand, 
$10^6$ of these machines had been produced, of which
$10^3$ were known to be faulty, then the inference would be
justified.

A formal argument helps to clarify the situation.  Let $H_1$,
$H_2$, $E$ be as defined in the last section.  Then the
conditional probabilities of the machine being not faulty or
faulty, given the observation
$E$,  can be calculated using Bayes' formula:
\begin{equation}
 P\left( H_r \left| E\right.\right)
=
\frac{
      P\left(E \left|  H_r \right.\right)
      P\left(H_r\right)
}{
      P\left(E \left|  H_1 \right.\right)
      P\left(H_1\right)+
       P\left(E \left|  H_2 \right.\right)
      P\left(H_2\right)
}
\label{eq:Bayes}
\end{equation}
On the first assumption $P(H_2)=1-P(H_1)=10^{-20}$,
implying 
$ P\left(H_2 \left|  E \right.\right)\approx 0$:  meaning
that the machine  is most unlikely to be faulty.   On
the second assumption $P(H_2)=1-P(H_1)=10^{-3}$, implying
$ P\left(H_2 \left|  E \right.\right)\approx 1$: meaning
that the machine is almost certain to be faulty.

Now this problem is unusual in that one might well  know the
proportion of faulty machines.  In a situation
like that Fisher~\cite{FisherStat} (pp.~8--39) has no objection
to the Bayesian methodology (he hardly could object: 
Eq.~(\ref{eq:Bayes}) is an elementary consequence of the basic
principles of probability theory).

Suppose, however, one does \emph{not} know 
the proportion of faulty
machines.  In that case, there are really only two
alternatives:   \emph{either} give up, and refuse to make any
prediction, \emph{or} (not to put too fine a point on it)
take a guess.

It does not have to be a \emph{blind} guess.  We are told
 that  $P\left(E \left|  H_2 \right.\right)/P\left(E \left| 
H_1 \right.\right)=10^{10}$.  This means that the tipping
point---the place where $P\left( H_2 \left|
E\right.\right)$  switches from virtually certain to highly
improbable---occurs at $P(H_2) \sim 10^{-10}$.  Our
background knowledge informs us that the total number of
machines is unlikely to exceed the total population of the
Earth.  We also know that there was a whole batch of faulty
machines---implying that the number of faulty machines is
$\gg 1$.   It follows that $P(H_2)$  is likely to be
$\gg 10^{-10}$:  implying  it is a fairly safe bet that
the machine is faulty.

Still, guessing clearly is involved, and that was
Fisher's~\cite{FisherStat} (pp.~8--39) reason for rejecting the
Bayesian approach (except in special cases).  Fisher, like most
$20^{\mathrm{th}}$ century statisticians, thought that guessing
has no place in science.  He and others therefore tried to
develop an alternative approach, in which conclusions would be
based
\emph{purely} on the actual data, without \emph{any}
 dependence on the
statistician's preconceived and (to a degree) arbitrary
notions.  They tried, in other words, to make statistics
purely objective.

We may sympathize with the intention.  No one sensible would
choose to rely on guesswork, when something better is
possible.  The trouble is that something better is \emph{not}
possible.  The orthodox methodology, which Fisher and others
developed, is not really any more objective than Bayesian
statistics.  It only appears to be more objective. 

Consider, for instance, the example discussed on pp.~68--78 of
Fisher~\cite{FisherStat}, where  a coin has come up heads on $3$
out of
$14$ tosses, and one wants to infer the probability of heads.  
Let $E$ be
the be the particular sequence which is observed, and let
$H_p$ be the hypothesis ``tosses independent and probability
of heads is $p$ on every toss''.   On the assumption that one
of these hypotheses must be true Bayes' formula gives
\begin{equation}
P\left( H_p \left|
E\right.\right) = K p^3 (1-p)^{11} P(H_p)
\label{eq:BayesFisher}
\end{equation}
where $K = \left(\int_{0}^{1} p^3 (1-p)^{11} P(H_p) dp 
\right)^{-1}$ 
  is a normalization constant, and $P(H_p)$,  
$P\left( H_p \left|
E\right.\right)$  are probability
densities. 
 Fisher does not want to use this formula as it
stands because the fact that we have to guess the function 
 $P(H_p)$
means that the conclusion will be contaminated with 
subjective assumptions.  He discusses two ways of trying to
get round that difficulty.

His preferred solution is the  method which I
described in the last section.  In Eq.~(\ref{eq:BayesFisher})
he deletes the subjective element represented by $P(H_p)$,
retaining only  the so-called likelihood
\begin{equation}
 P\left(E\left| H_p\right. \right)=p^3 (1-p)^{11}
\end{equation}
His grounds are that the likelihood ``represents that part of
Bayes' calculation provided by the data themselves''
(Fisher~\cite{FisherStat}, p.~72).  He then argues that the
smaller the likelihood, the less ``plausible'' the
corresponding value of
$p$.  More specifically he maintains  that values of $p$ for
which 
$P\left(E\left| H_p\right.
\right)\le (1/15)\times P\left(E\left| H_{3/14}\right.
\right)$ ---\emph{i.e.}
values of $p$ outside the interval $(0.04,0.52)$---are
``obviously open to grave suspicion'' (though he neglects to
say 
 \emph{why}
we should be suspicious; in particular, he fails
to explain what  is so special about the number
$15$).

Now it seems to me that, however Fisher may  
choose to verbally express it, he is here effectively working
on  the same assumption as Laplace.  If we follow
Laplace, and set
$P(H_p)=1$, then Eq.~(\ref{eq:BayesFisher}) becomes
$P(H_p|E)=K P(E|H_p)$.  So Fisher's likelihood is
proportional to Laplace's probability density. 
This means Fisher would say that $p_1$ is ``less
plausible'' than $p_2$ in exactly those cases where Laplace
would say that $p_1$ is ``less probable'' than $p_2$, and
not in any other case.  It therefore seems to me that
Fisher's ``less plausible'' is operationally equivalent to
Laplace's ``less probable''.   Again, Fisher says that values
outside the interval $(0.04,0.52)$ are ``open to grave
suspicion''.  More generally, he thinks that if $P\left(E\left|
H_p\right.
\right)\ll P\left(E\left| H_{3/14}\right.
\right)$ then $p$ is much less plausible than  $3/14$.  I find
it difficult to see any operational difference between this and
Laplace's belief, that  $p$ is much less
\emph{probable} than $3/14$.

At any rate, it is hard to see very much operational
difference.  It is, however, noticeable that, whereas
Laplace assigns an exact, numerical value to the probability
that
$p\in (0.04,0.52)$ (to be specific, he thinks the
probability  
$=0.98\dots$, to infinitely many decimals), Fisher
does not  commit himself to more than the vague qualitative
statement, that values of $p\notin (0.04,0.52)$ are ``open
to grave suspicion''.  Although Fisher \emph{orders} his
plausibilities, he generally tries to avoid giving them any
exact \emph{numerical} significance (apart from that
mysterious factor
$1/15$).   So perhaps the real motive for Fisher's
reworking  of the Bayesian argument is just the feeling
that, in a case like the present, exact quantification is
inappropriate.

Given Fisher's assumptions that would be a very reasonable
position.  As Laplace sees it probabilities have a purely
epistemic significance.  For him, the assignment $P(H_p)=1$ 
expresses the normative  principle that, in a case
where prior information is lacking, the logically correct
attitude is one of perfect indifference between the competing
alternatives.  Fisher, however, thinks  of
$P(H_p)$ as an actually existent quantity.  If one looks at
it from that point of view Laplace's statement, that
$P(H_p)=1$ \emph{precisely},  would be inappropriate even if
one had very detailed prior information.  In a case like the
present the most that would be appropriate is an order of
magnitude estimate.

If words are used in the ordinary sense (and Fisher says
nothing to indicate that they are being used in any other
sense) it is just not logically possible to believe that
something is not improbable whilst simultaneously viewing it
with grave suspicion.   So if what Fisher says is 
intelligible at all,  his claim must be that
$p\in
(0.04,0.52)$ with 
high probability.
The 
only\footnote{
       It is sometimes said (though not by Fisher) that one
      cannot
     meaningfully talk of the probability that $p\in
     (0.04,0.52)$ because $p$ is a ``parameter'', not a
     ``random
     variable''.  This position is only open to someonw who
     is
     prepared to accept that one cannot meaningfully talk of
    the
     probability that a randomly selected radioactive
     nucleus has
     half-life $>t$, or the probabiity that a randomly
    selected
     person has core body temperature $>\theta$.
}
 reason
he does not want to use the 
word ``probability'' is, I suggest, that this
word has connotations of numerical exactitude which words
such as ``likely'', ``plausible'', ``suspicious'' or
``confident'' lack.

Fisher would convey his meaning 
more clearly if, instead of scouring the dictionary for
synonyms (``likelihood'', ``plausibility'', \emph{etc.}), he
were simply
 to say that, on the \emph{assumption} that  $P(H_p)$ is
roughly constant, the \emph{probability} that $p\notin
(0.04,0.52)$  is of order
$10^{-2}$.  Of course, it would then be apparent that he is
relying on an assumption which he did not derive from the
data:  that he is, in other words, relying on guesswork (not
necessarily a \emph{blind} guess, but a guess nonetheless).

Fisher also discusses the confidence interval method. 
This would, for many, be the method of choice.  Fisher,
however, considers it inferior to the likelihood method just
described (correctly, as it seems to me).

In the case supposed $(0.03,0.56)$ is a $98\%$ confidence
interval for $p$ (the interval being constructed on the
principle that the probability of $3$ or fewer heads is
$\le 0.01$ if $p\ge 0.56$, and the probability of $3$ or more
heads is $\le 0.01$ if $p \le 0.03$).   The usual
justification for this is that one would, in the long run,
expect a $98\%$ confidence interal to cover the true value
of $p$ more than $98\%$ of the time (not exactly $98\% $ of
the time due to the  fact that the random variable is
discrete).

The  problem with this argument is that it relies on the
principle ``highly improbable = FAPP impossible'' which I
criticized in the last section.  Let $I$ be the 
interval  actually obtained.   The argument  is
that we should reject values of $p \notin I$ because
$P(I|H_p)$ is then $\le 0.02$, and $0.02$ is a small number. 
This is wrong twice over.  It is  wrong in the first place
because it attaches significance to the \emph{absolute}
value of 
$P(I|H_p)$ when, as we saw in the last section, it is only
the \emph{relative} values that are relevant (Fisher's
likelihood argument is in that respect preferable).   It is
wrong in the second place because  even a relatively larger
value of
$P(I|H_p)$ only translates into a greater probability for 
$H_p$ if we assume   $P(H_p)=1$  (because it is only then
that
$P(H_p|I)\propto P(I |H_p)$).

The idea that one should trust the confidence interval
approach because the expected  failure rate is
small has a strong hold on the orthodox imagination.  So let
us look at it from another angle.  Suppose we have a
balance which is guaranteed to be $99.9999\%$ reliable. 
Under most circumstances we will trust its readings. 
Suppose, however, we put a mosquito on the pan and the
instrument reads $1\; \mathrm{kg}$.  Then we will conclude that
the instrument is misreading---even though the probability of
that happening is $10^{-6}$.   Our strong prior conviction
that a tiny little insect cannot possibly have mass $1\;
\mathrm{kg}$ will outweigh our conviction, that the
instrument is most unlikely to deceive us.

Similarly here:  if we had a sufficiently strong prior
conviction that a coin is most unlikely to be biased, then
not even a $99.9999\%$ confidence interval would be enough
to persuade us otherwise.  If in practice a $99\%$ confidence
interval would usually induce a change of mind, that is
because we would not usually  have a strong initial
prejudice in favour of the coin being fair:  because, in
other words, we would usually work on the tacit assumption
that  $P(H_p)$ is more or less uniform.

Let us return to the question I posed at the beginning of
this section:  how many heads in succession should it take to
convince us that a coin is biased?  The answer is:  it
depends on our starting assumptions, as represented by the
function $P(H_p)$.  If you start out on the assumption that
the odds are
$10^{10^6}:1$ against the coin being biased then you will
retain your belief, that the coin is almost certainly  fair,
even though it has come up heads on each of the last million
tosses.  And you will be  right to do
so---\emph{given} your assumptions.  

Suppose  you take it to be a \emph{given fact}
that the coin is fair (as in elementary textbook problems). 
Suppose, in other words, that you start on the assumption
that $P(H_p)=\delta(p-0.5)$. Then nothing at all will shake
your belief.  Nor should it,  as elementary textbooks all
correctly say (the gambler's fallacy really is a
fallacy).

Of course, if someone did in fact persist in believing that
a lottery  is fair, even though the same person had won it
every week for the last $10$ years, we would mostly consider
their belief perverse, not to say irrational.  Yet their
confidence would be entirely justified, if their assumptions
were valid.   Any irrationality there may be is in those
starting assumptions.  Not in a subsequent misapplication of
the rules of probability.

The question is:  what are the \emph{right} starting
assumptions? How does one decide?  We would mostly consider
the assumption
$P(H_p) \sim 1$ appropriate in the case of a coin.  At any
rate, it is the assumption  which all orthodox
statisticians and most Bayesians do in fact make.  On the
other hand the assumption   $P(H_p) =\delta(p-0.5)$ would
strike most of us as quite unreasonable (when applied to
the real world; not, of course, when applied to an
imaginary textbook world).  What is the basis  for that
belief?  

This is, in essence, the problem of induction.  Consider the
question I posed in Section~\ref{sec:Introduction}:  how many
observations does it take to justify NASA's belief, that
gravity falls off as $1/r^2$?  Clearly,  a single
measurement would not be sufficient.  Nor, I think, ten (I
think this is a case where a $99\%$ confidence interval
would fail to convince).   But as the data keeps
coming in, and the hypothesis is each time confirmed, there
eventually comes a tipping point:  a place where, rightly or
wrongly (and it may be wrongly), our attitude changes from
 reserve to qualified assent.  The significance of the
function $P(H_p)$ is that it sets  the tipping
point for the coin tossing example.  The question is, in both
cases:  exactly \emph{where} do we set the tipping point?

This problem,  in one
form or another (usually in a much more complicated, subtle
and interesting form), occurs in every situation where one
needs to reach conclusions on the basis of limited
information.   When, exactly, does the evidence in support
of a proposition become so strong that one would be willing
to stake one's life on it? (as the Apollo astronauts staked
theirs on the approximate truth of the Newtonian law of
gravity).  The answer we give to this question is partly
definitive of scientific rationality.  So it cannot be
shirked.  At least, it cannot be shirked if we want to
have a standard of scientific rationality.

Nevertheless, people often try to shirk it.  There are two
reasons for this.  In the first place, the question  cannot
be settled experimentally.  It concerns the standard of
empirical evidence, and for that very reason it is itself
beyond  the reach of empirical evidence.   So we have to
fall back on our intuitive judgment.  The trouble with that
is that what strikes us as intuitively reasonable is 
likely to depend on the way our brains are wired (not to
mention possible educational and cultural influences).  An
alien being might intuit differently.

The other difficulty is that the choice of tipping point is,
to an extent, arbitrary.   Faced with the question ``Just
how many observations are needed before it is reasonable to
stake one's life on the approximate truth of a scientific
hypothesis?'' different individuals will make different
decisions.  There does not seem to be any basis for
singling out one of these as the unequivocally
\emph{correct} decision.  Similarly with the coin-tossing
example.  On intuitive grounds we would mostly reject
the assumption $P(H_p) = \delta(p-0.318)$ without
hesitation.  But, within certain  limits, one choice of
$P(H_p)$ seems as good as another.  

These features of probability in general, and induction in
particular, worried Newton, as they worried Hume,  and as
they have worried numerous others since.  They worry me. 
However, the problem seems insurmauntable.  Probabilistic
reasoning, and therefore science, does partly depend on
intuition and guesswork.  It is a fact of life.

\section{Single-Case Probabilities}
Until now I have been looking at what may be called
retrodictive probabilistic reasoning:  the case where one
argues back, from  observations already performed, to the
underlying probability distribution.  I now want to look at
the predictive case, where one argues in the opposite
direction, from a given probability distribution to
observations not yet made.

So let us ask:  what is the predictive content of the
statement ``event $X$ has probability $p$''?   The usual,
frequentist answer to this question is that the statement
has no predictive implications for the outcome of any
\emph{single} trial, but that it does have predictive
implications for the outcome of a \emph{long sequence} 
of trials.  Suppose, for instance, that a fair coin is
tossed
$10^4$ times.  Then, on the assumption that the tosses are
independent, the probability that the relative frequency of
heads will be outside the interval $(0.48,0.52)$ is $\sim
10^{-6}$.  So, although we cannot say anything useful about
a single coin toss, we can be nearly certain that in $10^4$
tosses the relative frequency of heads will be close to
$0.5$.

At first sight this account of the matter may seem
 convincing.  But if one looks a little more closely it
will be seen that the argument is, in fact, making a tacit
appeal to the concept of a single-case probability.  Let $Y$
be the event ``relative frequency of heads $\notin
(0.48,0.52)$''.  The argument is relying on the idea that,
because the probability of $Y$ is $\sim 10^{-6}$, therefore
it is a \emph{safe bet} that $Y$ will not occur in a
\emph{single} run of $10^4$ tosses.   This is not really any
different from arguing,  in respect of a \emph{single}
lottery draw, that because the probability of Alice winning 
is only
$10^{-6}$, therefore  we can be nearly certain
that Alice is not going to win.   The conclusion
seems reasonable enough, if one judges by the standards of
commonsense.  But it represents a clear departure from the
frequentist principle, that it is not possible to make any
valid probabilistic prediction regarding the outcome of a
single trial.

Every probability is a single-case probability at the point
of empirical application. Suppose, for example, it is known
that a certain operation has probability
$p$ of causing serious, irreversible brain damage.  The
question the patient has to decide is:  having in view all
the circumstances, is s/he willing to take that risk? 
Whichever way the patient decides, the decision will be
based on $p$ regarded as a single-case probability.

The decision is, in fact, a kind of bet.  The unpleasant
truth, which frequentists would prefer not to see, is that
probability has essentially to do with gambling.  It need not
be frivolous, or irrational gambling, as occurs in a
casino.  It may  be gambling in deadly earnest (as
in the above example).  But it is gambling nonetheless.  At
the point of empirical application every piece of predictive
probabilistic reasoning presents us with a dilemma of the
following general form ``Given that the probability of $X$
is $p$ are we, or are we not prepared to bet that $X$ will
in fact happen, in a single trial?''

If we are concerned with the outcome of a coin-tossing
experiment, and if $X$ is the event ``relative frequency of
heads $\in (0.48, 0.52)$'' then, by choosing a long enough
sequence, we can make the probability of $X$ as close to
$1$ as we wish.  But we cannot make it strictly $=1$.  So,
if we want to come to empirical conclusions, our only option
is to make a bet.  The bet may, we think, be very very safe. 
But it is still a bet.

Making the best decision in the face of
uncertainty---calculating the best bet---is what probability
is \emph{for}.  However distasteful it may be to
objectivist-minded philosophers, gambling is in fact the
point.  Remove the gambling element---remove the concept of
a single-case probability---and you remove with it all the
empirical applications.  What remains is not really
probability at all, but abstract measure theory.

To understand the content of probability statements one
needs to look at the point where probability collides with
reality.  One needs, in other words, to consider 
single-case probabilities.  When one does that it becomes
clear that a probability statement is, broadly speaking, a
statement about what we can reasonably expect.

Consider, once again, the case where Alice buys one ticket
in a lottery having $10^6$ tickets, and her ticket wins.  
Even after it is known that Alice \emph{did} win the
lottery, we would still say that Alice was very 
\emph{unlikely} to
win.  And we would be right to say it:  because the
statement, that Alice is unlikely to win, is not, primarily,
a statement about the actual outcome.  Rather, it is a
statement about what Alice, and us, could reasonably expect
regarding the outcome.  The fact, that Alice did win, does
not alter the fact, that she could not reasonably have
expected to win.

Probability, in short, is epistemic.

\subsubsection*{Acknowledgements}  I am grateful to C.A.~Fuchs,
 who first made me see the
importance of these questions, and also to
H.~Brown, P.~Busch, J.~Butterfield,  M.~Donald,  L.~Hardy,
T.~Konrad, P.~Morgan, R.~Schack, C.~Timpson and  J.~Uffink for
extremely stimulating discussions.


\begin{thebibliography}{99}
\bibitem{FuchsSam} C.A.~Fuchs,
e-prints quant-ph/0105039; quant-ph/0205039.
\bibitem{FuchsEtAl} C.M.~Caves, C.A.~Fuchs, and R.~Schack,
\emph{Phys.~Rev.~A}\ \textbf{65}, 022305 (2002);
\textbf{66}, 062111 (2002);
\emph{J.~Math.~Phys.}\ \textbf{43}, 4537 (2002).
\bibitem{FuchsSchack} C.A.~Fuchs and R.~Schack, e-print
quant-ph/0404156, to appear in \emph{Quantum Estimation
Theory}, edited by M.G.A.~Paris and J.~Rehacek
(Springer-Verlag, Berlin, 2004).
\bibitem{Appleby} D.M.~Appleby, e-print quant-ph/0402015.
\bibitem{GilliesA}  D.~Gillies, \emph{Philosophical Theories
of Probability} (Routledge, London, 2000).
\bibitem{Hald} A.~Hald, \emph{A History of Mathematical
Statistics from 1750 to 1930}  (Wiley, New York, 1998)
\bibitem{SklarA} L.~Sklar, \emph{Physics and Chance}
(Cambridge University Press, Cambridge, 1993).
\bibitem{vonPlato} J.~von Plato, \emph{Creating Modern
Probability} (Cambridge University Press, Cambridge, 1994).
\bibitem{Laplace} P.S.~de Laplace (trans. F.W.~Truscott and
F.L.~Emory),
\emph{A Philosophical Essay on Probabilities}  (Dover, New
York, 1951). French original published 1820.
\bibitem{JaynesA}  E.T.~Jaynes, \emph{Probability Theory: 
The Logic of Science} (Cambridge University Press,
Cambridge, 2003).
\bibitem{JaynesB} E.T.~Jaynes (ed. R.D.~Rosenkrantz),
\emph{Papers on Probability, Statistics and Statistical
Physics} (Reidel, Dordrecht, 1983).
\bibitem{JeffreysB} H.~Jeffreys, \emph{Theory of
Probability},
$3^{\mathrm{rd}}$ edition  (Clarendon Press, Oxford, 1961).
\bibitem{JeffreysC} H.~Jeffreys, \emph{Scientific Inference},
$3^{\mathrm{rd}}$ edition (Cambridge University Press,
Cambridge, 1973).
\bibitem{Ramsey}  F.P.~Ramsey, ``Truth and Probability'',
reprinted in L.~Sklar (ed.),
 \emph{Probability and Confirmation}
(Garland Publishing, New York, 2000).  First published 1931.
\bibitem{Fin2} B.~de Finetti,  ``Probabilism'', English
Translation,
\emph{Erkenntnis} \textbf{31}, 169 (1989).  Italian original
published 1931.
\bibitem{Fin1} B.~de Finetti (trans.\ A.~Mach\'{i} and
A.~Smith),
\emph{Theory of Probability} (Wiley, New York, 1975). 
Italian original published 1971.
\bibitem{Savage} L.J.~Savage, \emph{The Foundations of
Statistics},
$2^{\mathrm{nd}}$ edition (Dover, New York, 1972).
\bibitem{Bernardo} J.M.~Bernardo and A.F.M.~Smith,
\emph{Bayesian Theory} (John Wiley and Sons, Chichester,
1994).
\bibitem{Howson} C.~Howson and P.~Urbach, \emph{Scientific
Reasoning:  the Bayesian Approach} (Open Court, La Salle,
1989).
\bibitem{Earman}  J.~Earman, \emph{Bayes or Bust? A Critical
Examination of Bayesian Confirmation Theory} (MIT Press,
Cambridge Mass, 1992).
\bibitem{Mises} R.~von Mises, \emph{Probability, Statistics
and Truth} (Dover, New York, 1981).  Reprint of
$2^{\mathrm{nd}}$ revised English edition,
 published 1957.
\bibitem{MisesB} R.~von Mises (ed.\ H.~Geiringer),
\emph{Mathematical Theory of Probability and Statistics}
(Academic Press, New York, 1964).
\bibitem{Reich} H.~Reichenbach, \emph{The Theory of
Probability} (University of California Press, Berkelely,
1971).
\bibitem{PopperB} K.R.~Popper, \emph{The Logic of Scientific
Discovery} (Hutchinson, London, 1959).
\bibitem{Fraassen} B.C.~van Fraassen, \emph{The Scientific
Image} (Clarendon Press, Oxford, 1980).
\bibitem{GilliesB} D.A.~Gillies, \emph{An Objective Theory of
Probability} (Methuen, London, 1973).
\bibitem{PopperC} K.R.~Popper, ``The Propensity
Interpretation of Probability'', \emph{Brit.\ J.\ Phil.\
Sci.}\ \textbf{10}, 25--42 (1959).
\bibitem{PopperD} K.R.~Popper, \emph{Realism and the Aim of
Science} (Hutchinson, London, 1983).  
\bibitem{Fisher}  R.A.~Fisher (ed. J.H.~Bennett),
\emph{Statistical Methods, Experimental Design, and
Scientific Inference} (Oxford University Press, Oxford,
1990).
\bibitem{HumeA} D.~Hume (ed.~L.A.~Selby-Bigge, revised
P.H.~Nidditch),
\emph{A Treatise of Human Nature}, $2^{\mathrm{nd}}$ edition
(Clarendon Press, Oxford, 1978).  Originally published
1739-40.
\bibitem{Newton} I.~Newton, \emph{Opticks}, based on the
$4^{\mathrm{th}}$ edition, 1730 (Dover, New York, 1952).
\bibitem{Stove} D.C.~Stove, \emph{Popper and After:  Four
Modern Irrationalists} (Pergamon Press, Oxford, 1982).
\bibitem{NewSmith} W.H.~Newton-Smith in \emph{Karl Popper: 
Philosophy and Problems}, edited by A.~O'Hear (Cambridge
University Press, Cambridge, 1995).
\bibitem{Jeffrey} R.C.~Jeffrey,  in
R.E.~Butts and J.~Hintikka (eds), \emph{Basic Problems in
Methodology and Linguistics: 
$5^{\mathrm{th}}$ International Congress of Logic,
Methodology, and Philosophy of Science pt.~3} (Reidel,
Dordrecht, 1977).
\bibitem{FisherStat} R.A.~Fisher, \emph{Statistical Methods and
Scientific Inference}, $3^{\mathrm{rd}}$ edition.  Page
references to version reprinted in Fisher~\cite{Fisher}.
\end{thebibliography}
\end{document}